\documentclass{PoS}
\usepackage{subfigure}

\makeatletter
\newcommand{\figcaption}[1]{\def\@captype{figure}\caption{#1}}
\newcommand{\tblcaption}[1]{\def\@captype{table}\caption{#1}}
\makeatother

\def \mate<#1|#2|#3>{\mbox{$\langle {#1}|\,{#2}\,|{#3}\rangle$}}

\title{Baryon-baryon potentials in the flavor SU(3) limit from lattice QCD}

\ShortTitle{Baryon-baryon potentials in the flavor SU(3) limit from lattice QCD}

\author{\speaker{Takashi Inoue}\\
        Graduate School of Pure and Applied Sciences, University of Tsukuba,\\ 
        Tsukuba, Ibaraki 305-8577, Japan\\
        E-mail: \email{takash-i@sophia.ac.jp}}

\author{
for HAL QCD Collaboration
}
\author{
\begin{center}
\includegraphics[width=0.33\textwidth]{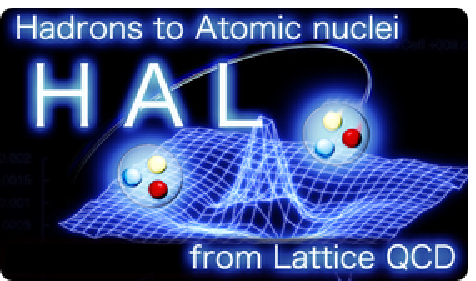}
\end{center}
}

\abstract{
We investigate baryon-baryon (BB) interactions in the flavor SU(3) symmetric world,
by using the 3-flavor full QCD simulations. In the SU(3) limit, 
six independent two-baryon states exist for a given orbital angular momentum.
We forcus on S-wave states and extract BB potentials using the method recently developed.
We discuss the flavor dependence of the potentials, 
in particular, the strength of the repulsion at short distance.
We find that the SU(3) flavor structure of the repulsive core is compatible 
with that expected from the Pauli principle for the quarks.
}

\FullConference{The XXVII International Symposium on Lattice Field Theory\\
		 July 26-31, 2009\\
		 Peking University, Beijing, China}

\begin{document}

\section{Introduction}
The generalized nuclear force, which includes not only the nucleon-nucleon (NN) interaction 
but also hyperon-nucleon (YN) and hyperon-hyperon (YY) interactions,
has been one of the central topics in hadron and nuclear physics.
Experimental studies on the ordinary and hyper nuclei as well as the 
observational studies of the neutron stars and supernova explosions are
 deeply related to the physics of the generalized nuclear force \cite{Hashimoto:2006aw}.

As far as the  NN interaction is concerned, significant number of
scattering data have been accumulated and are parametrized
in terms of the phase shifts or of the phenomenological
potentials  \cite{Machleidt:2000yw}. In particular, 
the moderate attraction at low energies and the strong repulsion 
at high energies seen in the S-wave NN scattering phase shift is one of the 
characteristic features of the NN force.

On the other hand, the properties of the YN and YY interactions are hardly known even today
due to the lack of high precision YN and YY scattering data.
The approximate flavor SU(3) symmetry does not necessarily create
a bridge between the NN force and the hyperon forces because
there are  six independent flavor-channels for the scatterings between octet baryons.
Although there have been previous theoretical attempts to fill the
gap on the basis of the phenomenological quark models  \cite{Oka:2000wj}
and  of the one-boson-exchange models \cite{Rijken:2001tw},
it is much more desirable to study the generalized nuclear force from first principle QCD. 
 
This report is our exploratory attempt to unravel the similarities and differences among different
flavor representations of the baryon-baryon (BB) potentials on the basis of the lattice QCD simulations. 
To capture the essential part of physics, we take the finite but degenerated u,d,s quark masses,
so that the system is in the flavor SU(3) limit.
In the next section, we briefly review our method to study the baryon-baryon interaction in lattice QCD. 
In Section 3, we introduce the classification of baryon-baryon systems in the flavor SU(3) limit.
In Section 4, we show a setup of our numerical simulations.
In Section 5, we show the baryon-baryon potentials in the SU(3) limit
and discuss their implications. The last section is devoted to summary and outlook.

\section{Baryon-baryon potentials from lattice QCD}

The potential is a useful concept to describe elastic BB interactions at low-energies.
Recently a method to extract the potential from lattice QCD simulations
has been proposed and applied to the NN system \cite{Ishii:2006ec,Aoki:2009ji}, 
and $\Xi N$ and $\Lambda N$ systems \cite{Nemura:2008sp,Nemura:2009kc,Nemura:2009lat} as well.
In this method, the potential is defined from the Bethe-Salpeter (BS) amplitude $\phi_E(\vec r)$ 
through the Schr\"odinger equation as
\begin{equation}
   \left[ \frac{\nabla^2}{2 \mu} + T \right] \phi_{E}(\vec r) 
 = \int\!\!d^3\vec r' \, U(\vec r, \vec r') \, \phi_{E}(\vec r'),
 \quad\mbox{with}\quad 
 T = \frac{k^2}{2\mu} ~,\quad
 k^2 = \frac{E^2}{4} - \frac{M^2}{4}
\end{equation}
where $\mu$ and $M$ are reduced and total masses of the system.
Here $U(\vec r, \vec r')$ is an energy independent but non-local potential.
At low energies, it can be expanded by the local velocity $\vec v = \vec{\nabla}/\mu$ as   
  $U(\vec r,\vec r') = V(\vec r, \vec v)\delta(\vec r-\vec r') 
  = (V_{LO}+ V_{NLO} + V_{NNLO} + \cdots )\delta(\vec r-\vec r') $,
where $N^nLO$ term is of $O(v^n)$.  Thus, in the leading-order (LO) approximation, we have
\begin{equation}
 V(\vec r) = \frac{1}{2\mu}\frac{\nabla^2 \phi_{E}(\vec r)}{\phi_{E}(\vec r)} + T ~.
 \label{eqn:vr}
\end{equation}
It is shown in Ref. \cite{Murano:2009} that the LO potential $V(\vec r)$ defined above
is a good approximation of $U(\vec r,\vec r')$ at low energies between $T \sim 0$ MeV and $T \sim 45$ MeV
for the $NN$ system in quenched QCD.
The BS wave function $\phi_E$ is extracted from the four point function, for example,
\begin{equation}
 W(t-t_{0},\vec r)
=\sum_{\vec x} \mate<0|p(t,\vec x+\vec r)\,n(t,\vec x)\,
                       {\bar n}(t_0)\,{\bar p}(t_0)|0> ~,
\end{equation}
which is dominated by the lowest energy state with the total energy $E_0$ at large time separation,
so that it is  proportional to $\phi_{E_0}(\vec r)$ at large $t-t_0$.

\section{Baryon-baryon potentials in the flavor SU(3) limit}

\begin{figure}[t]
 \begin{minipage}[c]{0.49\textwidth}
  \qquad \includegraphics[width=0.85\textwidth]{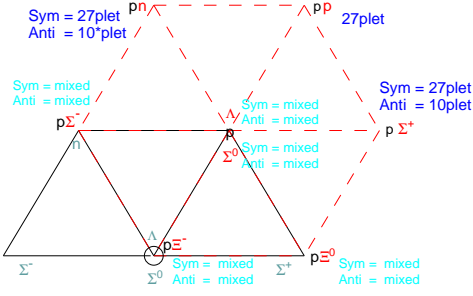}
  \caption{A part of all 64 octet-baryon pairs.
           Eight pairs composed 
           of the proton and octet baryons are expanded in several multiplets.}
  \label{fig:twobb}
 \end{minipage}
\hfill
 \begin{minipage}[c]{0.49\textwidth}
  \begin{center}
%  \begin{tabular}{lccc}
   \begin{tabular}{lcc}
    \hline
    \hline
    Hadron  & Mass [1/a] & Mass [MeV] \\
    \hline 
    $\pi$, $K$ &  0.5112(6) & ~834.4(1.0) \\
    $N$, $\Lambda$, $\Sigma$, $\Xi$ &  1.068(2)~\,  & 1744.0(3.8)\\
%   Hadron  & Mass [1/a] & Mass [MeV] & Fit range \\
%   \hline 
%   $\pi$, $K$ &  0.5112(6) & ~834.4(1.0) & 5-10\\
%   $N$, $\Lambda$, $\Sigma$, $\Xi$ &  1.068(2)~\,  & 1744.0(3.8)& 8-13\\
    \hline
    \hline
   \end{tabular}
  \end{center}
  \tblcaption{Hadron masses in the present lattice QCD simulation 
  ($16^3 \times 32$ lattice and spatial volume of about (2.0 fm)$^3$).
  The lattice spacing $a=0.1209$ fm is used to convert them in physical unit.
  }
  \label{tbl:mass}
 \end{minipage}
\end{figure}

In the flavor SU(3) limit, the ground states of a single baryon forms octet and decuplet.
Systems of two octet baryons  are classified into six irreducible multiplets in flavor SU(3); 
\begin{equation}
 8 \otimes 8 = \underbrace{1 \oplus 8 \oplus 27}_{\mbox{symmetric}} ~ 
            \oplus \underbrace{8 \oplus 10^* \oplus 10}_{\mbox{anti-symmetric}} \ \ .
\end{equation}
Since baryons obey the Fermi statistics, there exist six independent states
for a given orbital angular momentum. In particular for the S-wave,
we have singlet, octet and 27-plet for the spin-singlet state ($^1S_0$),
while octet, anti-decuplet and decuplet for the spin-triplet state ($^3S_1$).
The six independent potentials for these S-wave states are the targets of our study.
For example, we have computed $\sum_{\vec x} \mate<0|B_i\,B_j\, \overline{BB}^{(27)} |0>$
in order to extract $V^{(27)}$  through eq.(\ref{eqn:vr}),
where $\overline{BB}^{(27)}$ is a source operator of two octet baryons in the 27-plet representation,
and $B_i B_j$ is a sink operator  with two baryons which has an overlap with the 27-plet component.

We have to employ the source operator which belongs to the definite flavor representation 
in order to extract the corresponding BS wave functions at large time separation,
since all two octet baryon states have the same energy in the infinite volume and
possible energy differences due to the interactions in the finite volume are generally small.
Fig \ref{fig:twobb} illustrates a part of all octet-baryon pairs.
As shown in the figure, some of baryon pairs  belong exclusively to 27-plet, anti-decuplet or decuplet.
Numbers of wick contractions are relatively small for such operators,
while more complicated source operators with the SU(3) Clebsh-Gordan coefficients are needed
for other multiplets. % Therefore larger numbers of wick contractions are required. 
Due to the limitation of computational resources, we consider only the 
potentials $V^{(27)}$, $V^{(10^*)}$ and  $V^{(10)}$ for the moment,
which can be extracted by these simpler source operators.
 
\section{Numerical setup}
In order to calculate the potentials in 3-flavor full QCD,
we have utilized a gauge configuration set 
of Japan Lattice Data Grid(JLDG)/International Lattice Data Grid(ILDG)
generated by CP-PACS/JLQCD Collaborations on $16^3 \times 32$ lattice
with the RG improved Iwasaki gauge action at $\beta=1.83$ 
and the non-perturbatively O(a) improved Wilson quark action at lattice spacing $a= 0.129$ fm
and spatial volume of about (2.0 fm)$^3$ \cite{Ishikawa:2006ws,Ishikawa:2007nn}. 
The hopping parameters of the configuration set correspond
to the flavor SU(3) symmetric point, $\kappa_u=\kappa_d=\kappa_s=0.13760$.
Quark propagators are calculated from the spatial wall source at $t_0$
with the Dirichlet boundary condition in temporal direction at $t=t_0\pm 16$ mod 32.
The wall source is placed at 4 different time slices on each of 800 gauge configurations,
in order to enhance the signals, together with the average over forward and backward propagations in time. 
The average over discrete rotations of the cubic group is taken for the sink operator, 
in order to obtain $A_{1}^{+}$ state, whose lowest energy state is assumed to be dominated by the S-wave.
The statistical data are divided into bins of the size eight
and the jackknife prescription is adopted to estimate the statistical error.
All of our numerical computations are carried out at KEK supercomputer system, Blue Gene/L and SR11000.

\section{Results}

\begin{figure}[t]
 \begin{minipage}[t]{0.49\textwidth}
  \includegraphics[width=1.0\textwidth]{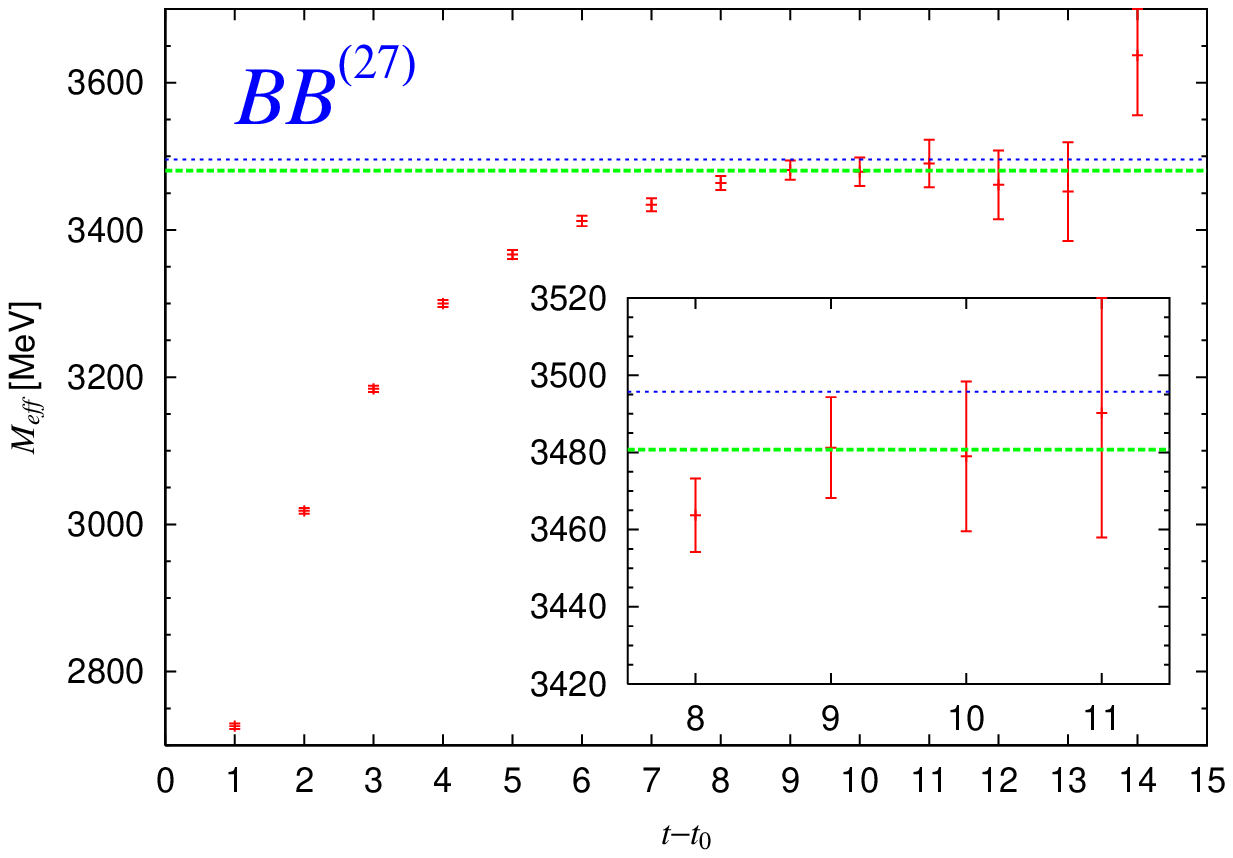}
  \caption{Effective mass plot of the 4-point function for the two-baryon operator in the 27-plet.
           The horizontal  lines stand for the threshold of two free baryons and its  error band. 
 }
  \label{fig:MeffBB}
 \end{minipage}
\hfill
 \begin{minipage}[t]{0.49\textwidth}
  \includegraphics[width=1.0\textwidth]{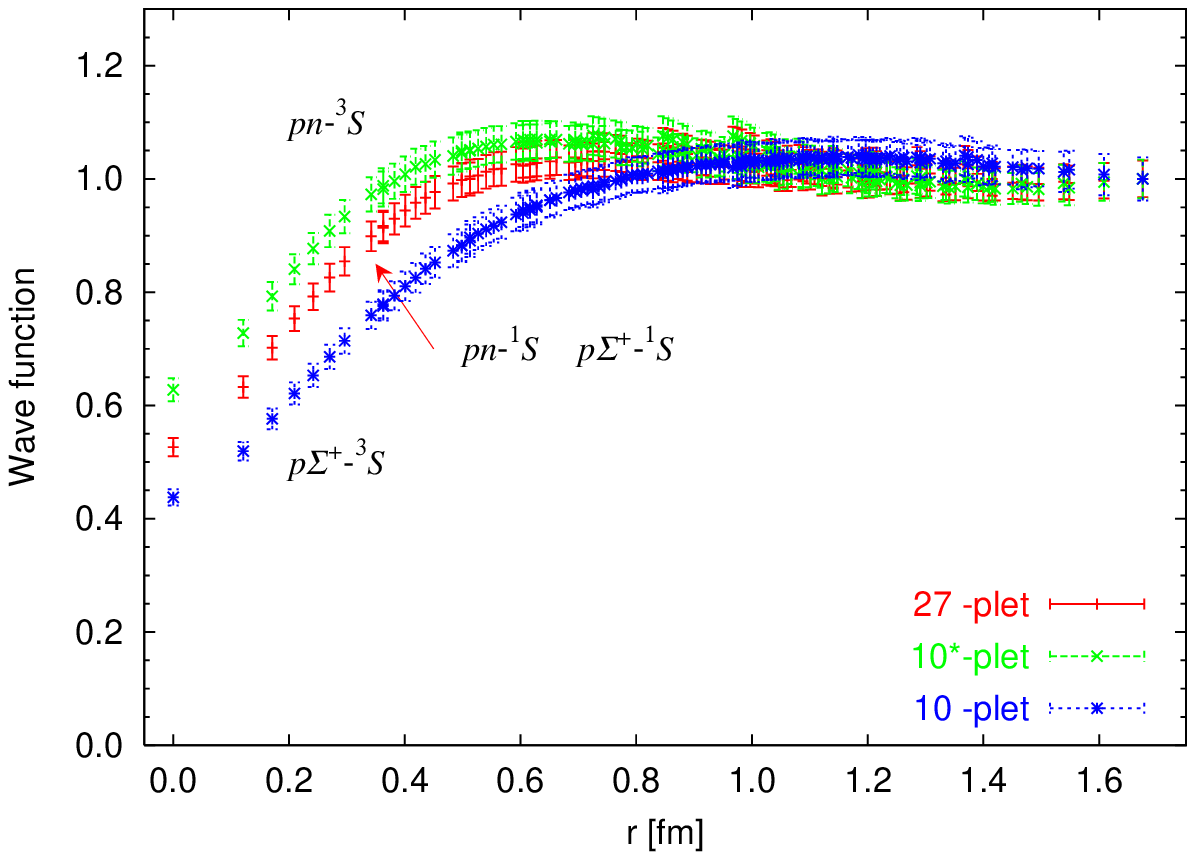}
  \caption{Two baryon wave functions in 27-plet (red), anti-decuplet (green) decuplet (blue)
           representations, normalized to one at the maximum distance.}
  \label{fig:bb-wave}
 \end{minipage}
\end{figure}

Table \ref{tbl:mass} lists masses of the octet pseudo-scalar meson and baryon in our simulation,
obtained by fitting corresponding 2-point functions.
%whose effective mass plots are given in Fig \ref{fig:Meff}.
The mass of the octet pseudo-scalar meson is 834 MeV,
which is much heavier than the physical $\pi$ or $K$ meson. 
For our present purpose, however, the flavor SU(3) symmetry is more important
than the lighter $u$, $d$ quark masses.

Fig \ref{fig:MeffBB} shows the effective mass plot of the 4-point function
for  two baryon operator in the 27-plet representation. 
A plateau is observed at large time separation, $t-t_0 \ge 9$, so that
the 4-point function is well saturated by the two-baryon ground state. 
A similar plateaus are obtained  for anti-decuplet and decuplet representation.
We take $t-t_0=10$ to extract baryon-baryon potentials in this report.

\begin{figure}[t]
 \begin{minipage}[t]{0.49\textwidth}
  \includegraphics[width=1.0\textwidth]{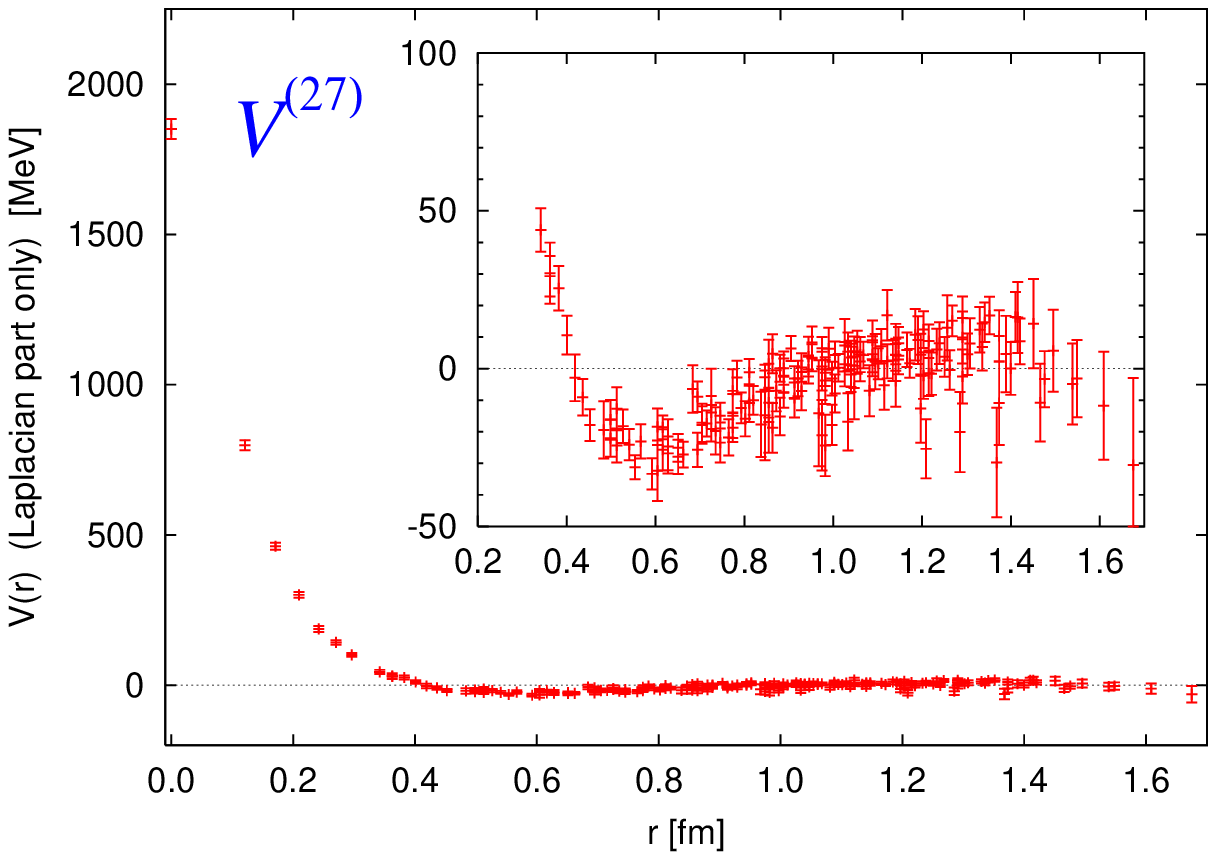}
  \caption{The 27-plet potential, $V^{(27)}(r)$, without energy shift $T_0$. }
  \label{fig:vr_27}
 \end{minipage}
\hfill
 \begin{minipage}[t]{0.49\textwidth}
  \includegraphics[width=1.0\textwidth]{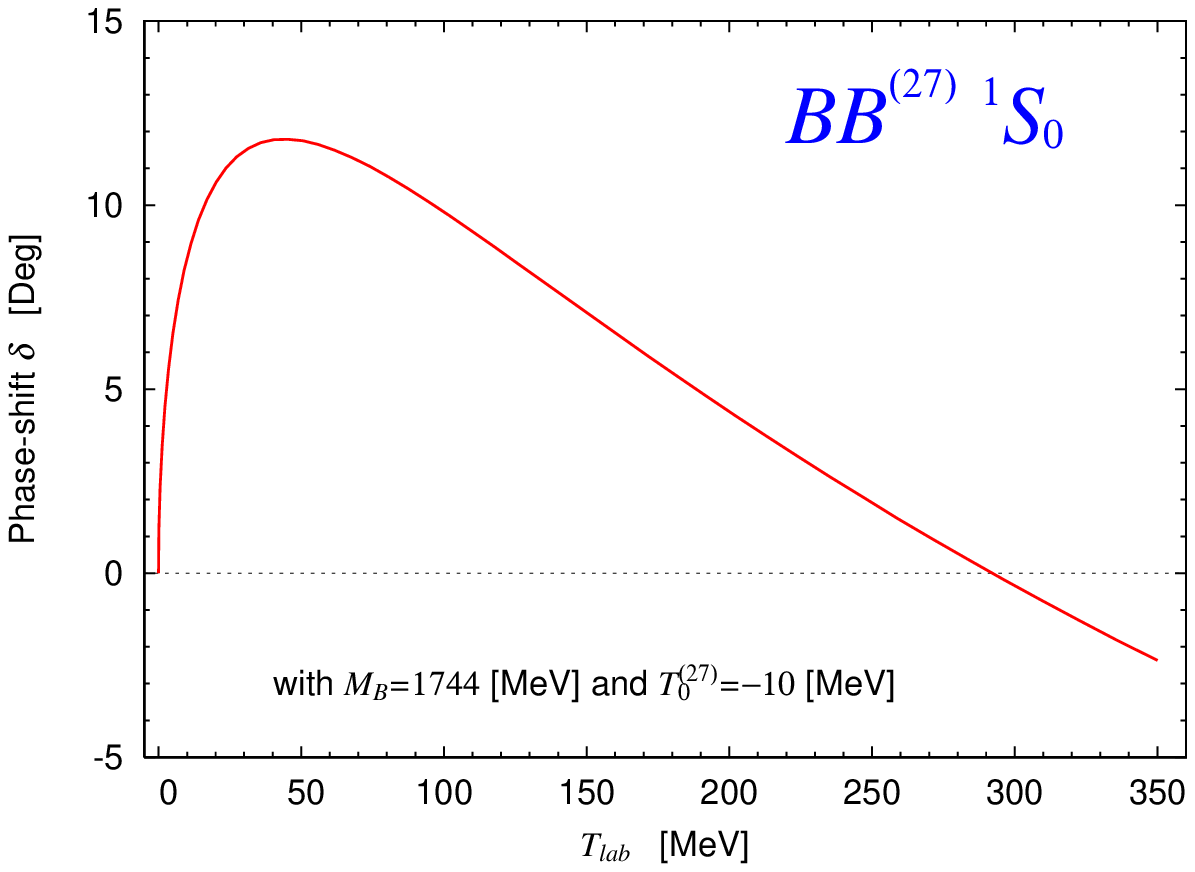}
  \caption{Scattering phase shift computed from the Schr\"odinger equation with $V^{(27)}$,
  as a function of the laboratory energy $T_{lab}$.}
  \label{fig:pnphase}
 \end{minipage}
\end{figure}

Fig \ref{fig:bb-wave} shows the BS wave function
obtained from lattice QCD simulations. 
It is clearly seen that the wave functions depend on their flavor representation(27, 10 or 10$^*$).
An agreement  between $pn$ $^1S_0$ state and $p\Sigma^+$ $^1S_0$ state,
both of which are the members of the 27-plet, gives a good consistency check of our calculations.
These wave functions suggest presences of repulsive interaction at short distance (around 
$r<0.4$ fm) for all channels.
% and attractive interaction at intermediate distance for 27-plet and anti-decuplet channels.
The shape of these wave functions indicates that no bound state seems to exist for all channels, though
more detailed analysis is needed in the finite volume for the definite conclusion.

In Fig \ref{fig:vr_27}, the potential extracted from the BS wave function
is plotted as a function of $r$ in the case of the 27-plet (spin singlet) state.
Note that here only the first term of eq.(\ref{eqn:vr})
is presented without the constant term $T_0$, which is expected to be small from Fig \ref{fig:MeffBB}.
The potential $V^{(27)}$ in Fig \ref{fig:vr_27} corresponds to
the NN $^1S_0$ potential in the SU(3) limit.
We see a repulsive core at short distance and an attractive pocket around 0.6 fm.
These qualitative features are similar to what was found in the NN system \cite{Ishii:2006ec}.
Assuming that the leading order term in the velocity expansion works well
as demonstrated for the NN  potentials \cite{Murano:2009},
we can calculate the ``physical'' observables in the SU(3) symmetric world. 
For example, Fig \ref{fig:pnphase} shows the scattering phase shift
as a function of energy in the laboratory system,
computed from the Schr\"odinger  equation with the baryon mass of 1744 MeV in the present simulation
and with the potential $V^{(27)}$ where $T_0= -10$ MeV is included.

\begin{figure}[t]
 \begin{minipage}[t]{0.49\textwidth}
  \includegraphics[width=1.0\textwidth]{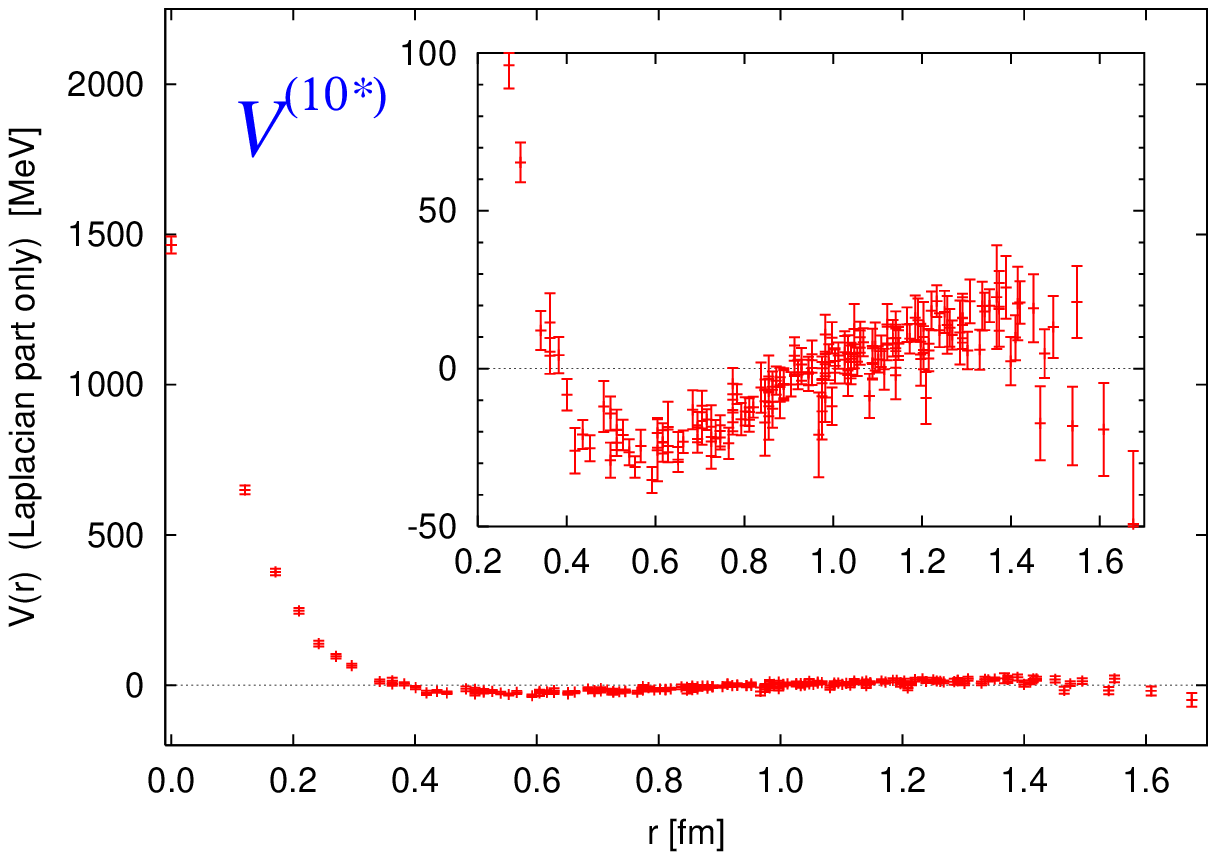}
  \caption{Potential $V^{(10^*)}(r)$ without $T_0$.}
  \label{fig:vr_10*}
 \end{minipage}
\hfill
 \begin{minipage}[t]{0.49\textwidth}
  \includegraphics[width=1.0\textwidth]{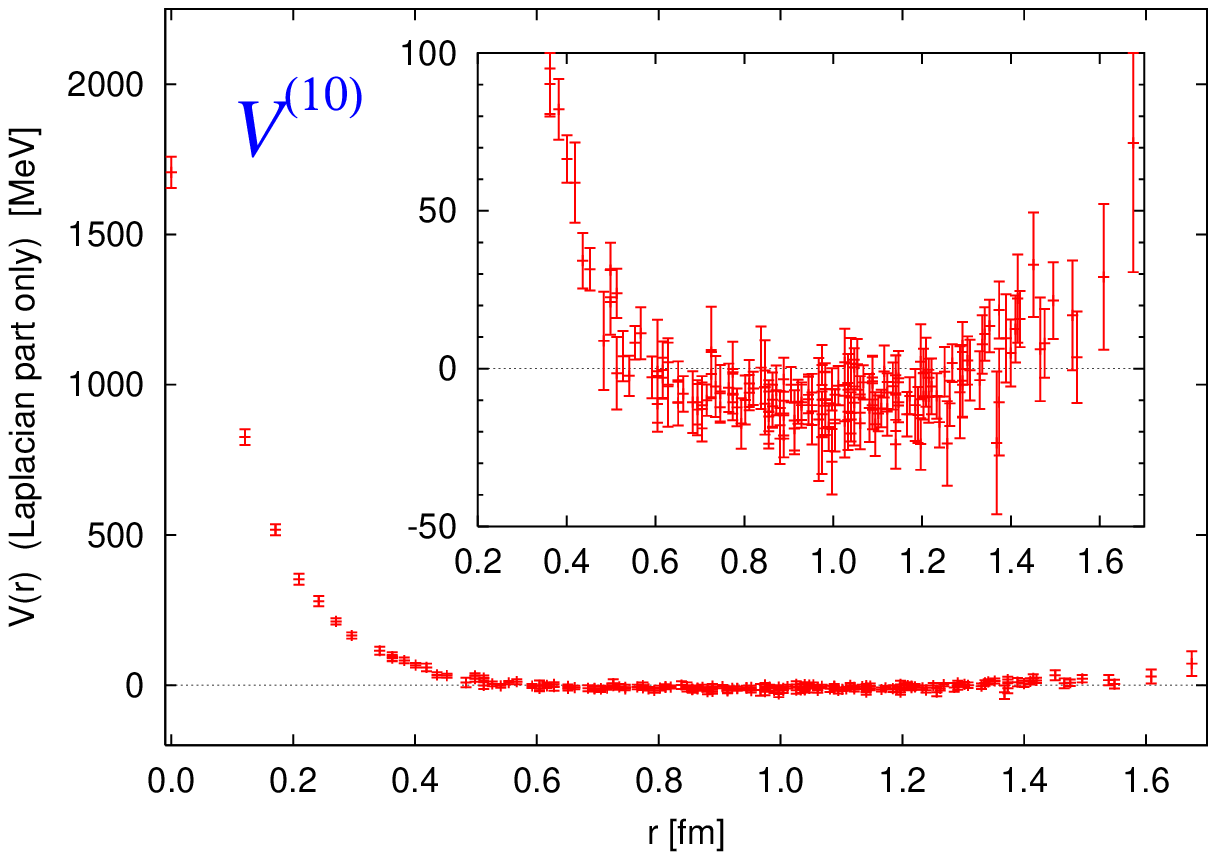}
  \caption{Potential $V^{(10)}(r)$ without $T_0$}
  \label{fig:vr_10}
 \end{minipage}
\end{figure}

The potential $V^{(10^*)}$ without $T_0$ in Fig \ref{fig:vr_10*} corresponds to
the NN $^3S_1$ effective central potential in the SU(3) limit.
We call it the "effective" central potential, since
the mixing with the $^3D_1$ state is not separated from the true central potential in this calculation.
We observe that $V^{(10^*)}$ has a weaker repulsive core and a broader attractive pocket
than $V^{(27)}$ which corresponds to the NN $^1S_0$ potential.
Accordingly, it is more attractive as a whole.
%Therefore
%it would be interesting to study this anti-decuplet potential 
%in lattice QCD at  lighter SU(3) symmetric point
%to see whether a bound state in the decuplet (``deuteron'')  exists or not.

\begin{figure}[t]
 \begin{minipage}[t]{0.49\textwidth}
  \includegraphics[width=1.0\textwidth]{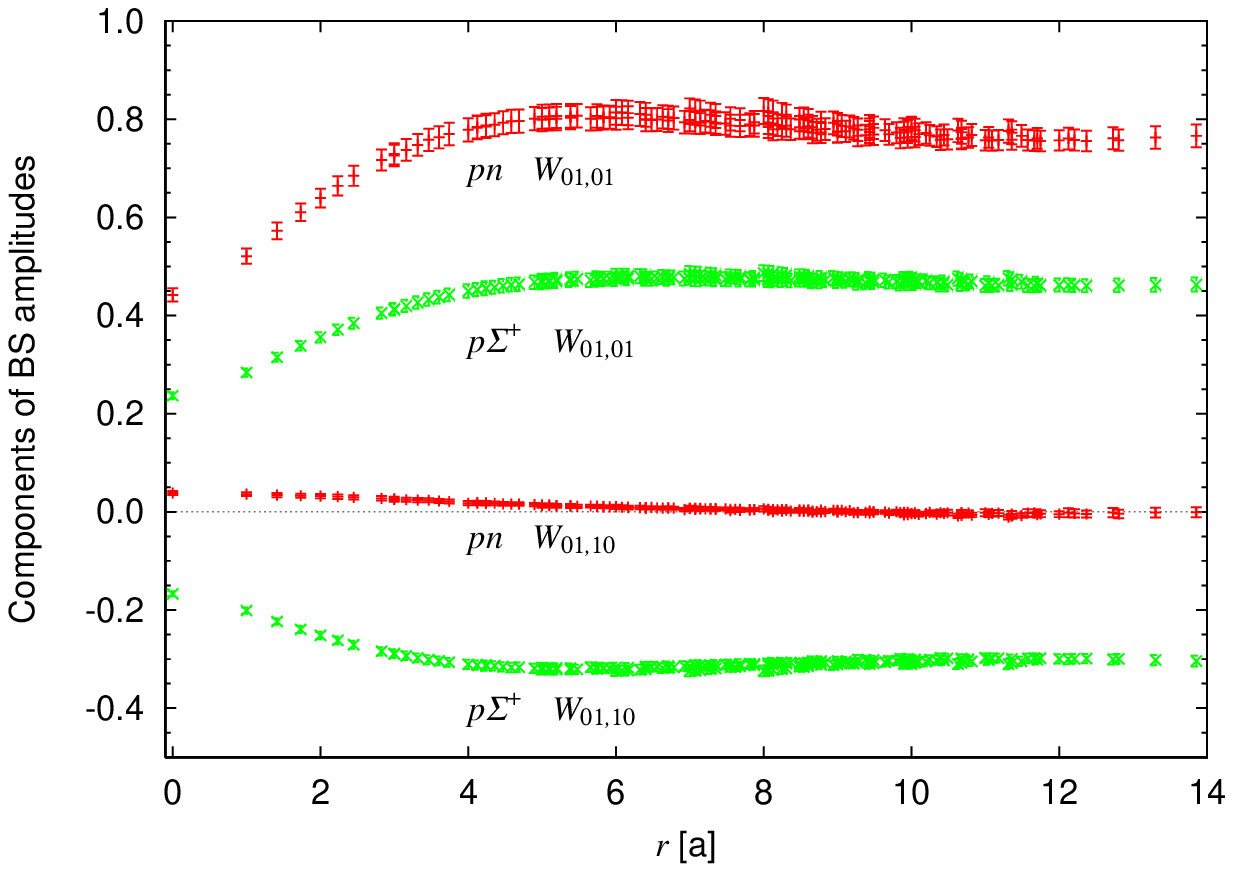}
  \caption{Component BS amplitudes with Dirac indexes of baryon operator. 
           Wave function of decuplet is given by sum of 
           $p\Sigma^+$ $W_{0101}$ and $p\Sigma^+$ $W_{0110}$.}
  \label{fig:partial}
 \end{minipage}
\hfill
 \begin{minipage}[t]{0.49\textwidth}
  \includegraphics[width=1.0\textwidth]{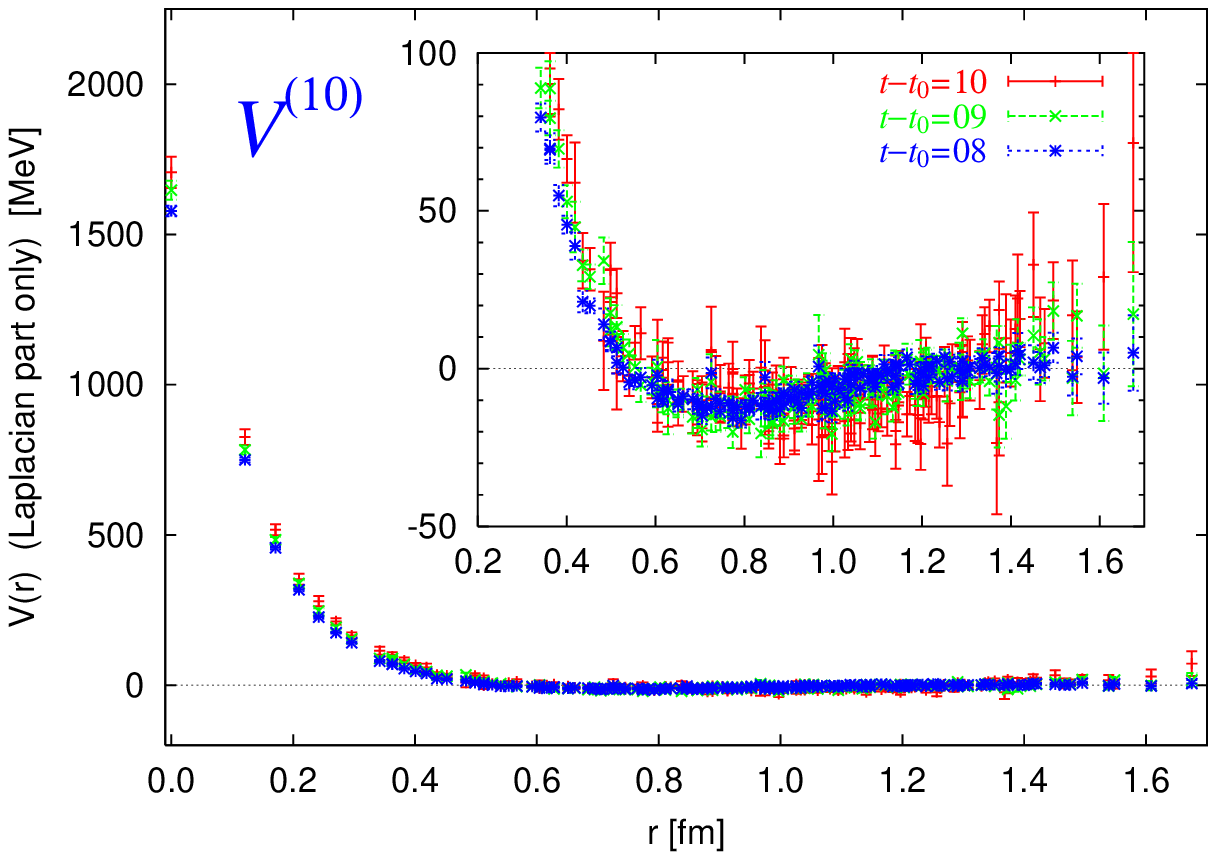}
  \caption{Potential $V^{(10)}(r)$ without $T_0$, extracted from data at three time slice.}
  \label{fig:v10sat}
 \end{minipage}
\end{figure}

The potential $V^{(10)}$ without $T_0$ in Fig \ref{fig:vr_10} looks different from others:
its repulsive core is higher and broader, and no clear attractive pocket is observed.
Accordingly, it is much more repulsive than others.
This feature can be also seen  from the wave function in Fig \ref{fig:bb-wave}.
The stronger repulsion in the decuplet channel has been predicted from the 
phenomenological quark model where the decuplet states are semi-forbidden state
due to the Fermi statistics of quarks \cite{Oka:2000wj}. 
%the most part of six quark state with identical quark orbital wave function,
%is excluded because of violation of the Fermi statistics (total anti-symmetry),
%and this this exclusion leads to the strong repulsion t short distance in this state
Our lattice QCD result is compatible with this Pauli principle picture.
This is more clearly seen in Fig \ref{fig:partial}, 
where the 4-point functions $W_{\alpha\beta,\alpha'\beta'}(r)$ 
with spinor indices $\alpha\beta,\alpha'\beta'$ are plotted.
The spin triplet (flavor anti-symmetric) component is given by the sum $W_{01,01}+W_{01,10}$. 
One can see that two $p\Sigma^+$ (decuplet) 4-point functions tend to cancel each other,
so that the decuplet wave function tends to vanish as if it were the totally forbidden state, 
while no such cancellation is observed for the anti-decuplet.
The potential $V^{(10)}$ likely has a shallow attractive pocket around 0.8 fm,
as seen more clearly in Fig \ref{fig:v10sat} at time slice $t-t_0=8$. 

\section{Summary and outlook}
We have performed the 3-flavor full QCD  simulation
to study the general features of the baryon-baryon interaction in the flavor SU(3) limit.
From the BS wave function measured on the lattice, 
we extracted the leading-order potentials in the S-wave, $V^{(27)}$, $V^{(10^*)}$ and $V^{(10)}$,
labeled by irreducible representations of flavor SU(3).
All three potentials have a repulsive core at short distance in common,
but its strength depends on the representation. 
In particular,  $V^{(10)}$ is most repulsive among three, which may indicate
the importance of the Pauli principle for quarks in baryon-baryon interactions at short distance.
Also, both $V^{(27)}$ and $V^{(10^*)}$ show  attractive pockets at intermediate distance.
We are currently working on the the potentials in the flavor singlet and octet channels
to obtain a complete picture of the BB interaction in the SU(3) limit. 

\section*{Acknowledgments}
We thank Columbia Physics System \cite{CPS} for their lattice QCD simulation code,
of which modified version is used in this work.
This work is supported by the Large Scale Simulation Program No.09-23(FY2009)
of High Energy Accelerator Research Organization(KEK). 
T.I. is supported in part by the Grant-in-Aid of the Ministry of Education,
Science and Technology, Sports and Culture (No. 20340047).

\end{document}